\documentclass[prb,showpacs,twocolumn,superscriptaddress,floatfix]{revtex4}
\usepackage[english]{babel}
\usepackage{graphicx}
\usepackage{amsmath}
\usepackage{amssymb}
\usepackage{epsfig}

\newcommand{\bsigma}{\boldsymbol{\sigma}}
\newcommand{\vm}{\mathbf{m}}

\begin{document}

\author{B. B{\'e}ri}
\affiliation{Instituut-Lorentz, Universiteit Leiden, P.O. Box 9506,
2300 RA Leiden, The Netherlands}

\author{J.\ N.\ Kupferschmidt}
\affiliation{Laboratory of Atomic and Solid State Physics, Cornell
  University, Ithaca, NY 14853-2501, USA}

\author{C.~W.~J.~Beenakker}
\affiliation{Instituut-Lorentz, Universiteit Leiden, P.O. Box 9506,
2300 RA Leiden, The Netherlands}

\author{P.\ W.\ Brouwer}
\affiliation{Laboratory of Atomic and Solid State Physics, Cornell
  University, Ithaca, NY 14853-2501, USA}

\date{November 2008}

\title{Quantum limit of the triplet proximity effect in
half-metal - superconductor junctions}
\begin{abstract}
We apply the scattering matrix approach to the triplet proximity effect in
superconductor-half metal structures. 
We find that for junctions that do not mix different orbital modes, 
the zero bias Andreev conductance vanishes, while the
zero bias Josephson current is nonzero. We illustrate this finding on a
ballistic half-metal--superconductor (HS) and
superconductor--half-metal--superconductor (SHS) junction with translation invariance
along the interfaces, and on HS and SHS systems 
where transport through the half-metallic region takes place through a single conducting channel. 
Our calculations for these physically single mode setups -- single mode point contacts and chaotic quantum dots with single mode contacts -- illustrate the main strength
of the scattering matrix approach: it allows for studying systems in the quantum mechanical limit, which is  
inaccessible for quasiclassical Green's function methods, the main 
theoretical tool in previous works on the
triplet proximity effect. 
\end{abstract}
\maketitle{}

\section{Introduction}

The recent experimental observation of the Josephson effect in a
half-metallic junction between two superconducting 
reservoirs\cite{keizer2006sts} has renewed interest in 
superconductor--ferromagnet hybrid devices. The observation of 
a supercurrent in a half metal is remarkable, because Cooper 
pairs in spin-singlet
superconductors consist of a pair of electrons with opposite spin,
whereas a half metal conducts electrons of one spin direction
only.\cite{groot83,pickett2001hmm,coey2002hmf} The resolution of this 
apparent paradox is the so-called
``triplet proximity effect'', first predicted theoretically
by Bergeret, Volkov, and Efetov.\cite{Berg01a}  (See also Ref.\
\onlinecite{Kadig01,BergeretPRB01,eschrig2003thm}, as well as Ref.~\onlinecite{bergeretRMP}
for a review.)  The triplet
proximity effect relies on the conversion of spin-singlet Cooper pairs 
of electrons with opposite spin into pairs of electrons
of equal spin at a spin-active interface between the superconductor
and the half metal.\cite{Berg01a,Kadig01,eschrig2003thm} Since pairs 
of equal-spin electrons can be transmitted coherently through a half
metal, the triplet proximity effect can indeed explain the observation of 
a Josephson current in the experiment.

Most theoretical studies of the triplet proximity effect were done
using the quasiclassical Green's function
method.\cite{Berg01a,Kadig01,BergeretPRB01,eschrig2003thm,Braude07,eschrig2007spc,eschrig2008tsc,Galak08,Asa07a,Asa07b}
This method is appropriate for
systems in which transport takes place through many conducting
channels.\cite{Eilenberger,LarkinOvchinn}   
For systems with few channels only, the Green's function
technique should be applied without the quasiclassical
approximation. This, albeit doable,\cite{Galak08,Asa07a,Asa07b} can lead to
calculations of significant complexity. 
Another method that is particularly well
suited for few channel structures is the scattering matrix
approach. This method has been frequently used in the context of
transport problems involving superconductors (for a review, see
Ref.~\onlinecite{RMTQTR}). However, it has not yet been applied to the
triplet proximity effect. It is the goal of the present article to
fill this gap.

In the language of the scattering approach, the triplet proximity effect
relies on the coherent Andreev reflection of electron-like excitations
into hole-like excitations with
the same spin.\cite{Qun08} 
Conventional Andreev reflection, as it takes place at
the interface between a normal metal and a superconductor, consists of
the reflection of an electron into a hole with opposite spin. 
  "Same spin" and "opposite spin" here refers to the spin band from which the electron and hole are taken. 
 Since electron and hole from the same spin band have opposite angular momentum, conservation of angular 
 momentum implies that electron and hole are from opposite spin bands. Hence, Andreev reflection of electrons 
 into holes from the same spin band requires that the interface
 between the half metal and the
superconductor is spin active. Examples of appropriate spin active 
interfaces are a thin ferromagnetic or half-metallic layer with a 
polarization that is non-collinear with the half metal's polarization
or a normal-metal spacer layer with strong spin-orbit scattering.

Our focus is on systems with the fewest number of channels possible,
a single conducting channel at the Fermi level. This limit can be 
achieved by having single channel contacts
between the superconductor(s) and the half metal. As an example of
this limit, we use the scattering theory to address the 
simplest single channel half-metal--superconductor (HS) junction 
that can display triplet proximity effect: 
a single channel ferromagnetic or half-metallic 
ballistic point contact between H and S electrodes. 
To study a more complex situation,  we investigate HS and 
superconductor--half-metal--superconductor (SHS)
junctions where the half metal is a chaotic quantum dot with 
single-channel point contacts. We also study the case of ballistic 
devices which have translation 
invariance along the interfaces. This situation allows for a single 
channel description as well, since the translation symmetry ensures
that  different transverse modes do 
not mix. While the latter system can in principle be 
addressed by the quasiclassical Green's function method, the
former, physically single channel setups are fully quantum mechanical, 
hence falling outside of the scope
of quasiclassics.

We use the scattering matrix approach to calculate the differential
conductance of an HS junction, and the (zero-bias) supercurrent in an
SHS junction.
We find that there is a remarkable difference between
these two observables in the single-channel limit. For a single-channel
half-metal--superconductor junction at zero temperature, the 
linear conductance vanishes
at the Fermi level. The conductance becomes appreciable only if the applied
voltage is comparable to the superconducting gap $\Delta$ or to
the Thouless energy of the junction, whichever is smaller. The
Josephson current, on the other hand, proves to be nonzero at zero
temperature. 
 The origin of this different behavior is that the
Josephson effect contains information about the entire excitation
spectrum of an SHS junction, whereas the linear conductance is a property that
requires knowledge of excitations at the Fermi-level only.

\begin{figure}
\epsfysize=0.5\hsize
\epsffile{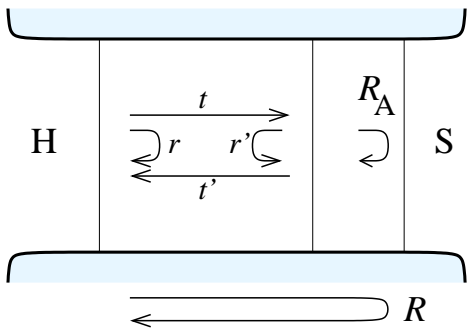}
\caption{\label{fig:1} (Color online) Composite
HS junction consisting of a half-metallic
  contact (left), a superconducting contact (right), and a
  spin-active
  intermediate layer (center). In most of our considerations, the
  intermediate layer is taken to be ferromagnetic with a magnetization
  direction not collinear with the polarization of the half metal.
Transport through the HS junction is described by the scattering
  matrix ${\cal R}$, which is calculated in terms of the Andreev
  reflection matrix ${\cal R}_{\rm A}$ of an ideal
  normal-metal--superconductor interface and the reflection and
  transmission matrices $r$, $r'$, $t$, and $t'$ of the non-superconducting region.}
\end{figure}

The remainder of this article is organized as follows. In Sec.\
\ref{sec:2} we outline the key elements of the scattering approach and
its application to HS junctions with a spin-active superconductor
interface. In Secs.\ \ref{sec:3} and \ref{sec:4} we then apply the
scattering theory to transport through an HS junction and to the
Josephson effect in an SHS junction, respectively. We conclude in Sec.\
\ref{sec:5}.

\section{Scattering approach}
\label{sec:2}

For a scattering description of the triplet proximity effect, we
consider half-metal--superconductor (HS) junctions that consist of a
half metal ``end'', a spin-active intermediate layer, and a
superconductor. The intermediate layer may be 
half-metallic, ferromagnetic, or normal metallic.

The central object in the scattering approach is the scattering matrix 
${\cal R}(\varepsilon)$ of the HS junction. It relates the amplitudes of
excitations at energy $\varepsilon > 0$
propagating towards the superconductor and excitations 
propagating away from the superconductor at the half-metal end of
the junction, see Fig.\ \ref{fig:1}. 
If $\varepsilon$ is below the superconducting gap $\Delta$, all
excitations must be reflected at the interface with the
superconductor. This reflection can be of normal type (electron-like
excitations are reflected as electrons, and hole-like excitations are
reflected as holes), or of Andreev type (electron-like excitations are
reflected as holes and vice versa). Both reflection types are 
contained in the matrix ${\cal R}$, which is made explicit by the
decomposition
\begin{equation}
{\cal R}(\varepsilon)=
\begin{pmatrix}
r_{\rm ee}(\varepsilon) & r_{\rm eh}(\varepsilon)\\
r_{\rm he}(\varepsilon) & r_{\rm hh}(\varepsilon)
\end{pmatrix},
  \label{eq:Reeeh}
\end{equation}
where $r_{\rm ee}$ and $r_{\rm hh}$ are matrices
that describe normal
reflection, whereas $r_{\rm eh}$ and $r_{\rm he}$ describe Andreev
reflection. All four matrices have dimension $N$, the number of 
propagating modes at
the Fermi level in H. Note that the propagating modes in H are not
spin degenerate. 
Below, we will use the polarization direction of H as the spin
quantization axis and refer to the electrons with spin parallel to 
the polarization direction of H as ``spin up''.

\begin{figure}
\epsfysize=0.5\hsize
\epsffile{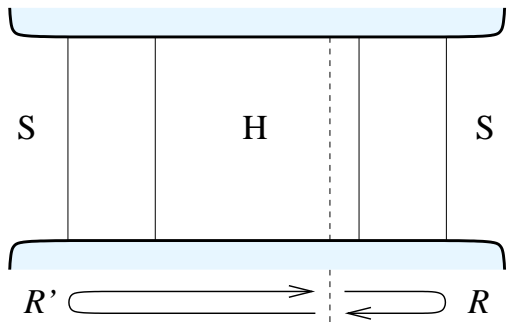}
\caption{\label{fig:2} 
(Color online) Schematic drawing of an SHS junction. In the scattering approach, an 
SHS junction is seen as two opposing (composite)
HS junctions, with scattering
matrices ${\cal R}'$ and ${\cal R}$, respectively. In the calculations
of Sec.\ \ref{sec:BallisticSHS}, scattering phase shifts from the central
half-metallic part are included into ${\cal R}'$.}
\end{figure}

Knowledge of the scattering matrix ${\cal R}$ is sufficient to
calculate the conductance of an HS junction, as well as the Josephson
current in an SHS junction. The zero temperature differential
conductance of an HS junction reads\cite{BTK82,takane1992cfm}
\begin{equation}
  G(eV)=\frac{2 e^2}{h} {\rm Tr}\ r_{\rm he}^\dagger(eV) 
  r_{\rm he}(eV).
\label{eq:GNSgen}
\end{equation}
(The factor of 2 accounts for the doubling of the current by the conversion of an electron into a hole.)
An SHS junction can be viewed as two HS junctions
opposed to each other, see Fig.\ \ref{fig:2}. Denoting the scattering
matrix corresponding to the second junction as ${\cal R'}$, the 
Josephson current reads\cite{brouwer1997}
\begin{equation}
  I=-\frac{2 e k_B T}{\hbar}
  \frac{d}{d\delta\phi}\sum_{n=0}^\infty \ln \det [1-{\cal
    R'}(i\omega_n){\cal R}(i\omega_n)],
\label{eq:Ijosgen}\end{equation}
where $\omega_n=(2n+1)\pi k_BT$ are the Matsubara frequencies, and $\delta
\phi$ is the phase difference between the two superconductors.

In principle, the explicit calculation of ${\cal R}$ requires a
solution of the Bogoliubov-de Gennes equation for the full HS
junction. Here, we take a different approach,\cite{Bee92a} and express ${\cal R}$
in terms of the scattering matrix ${\cal S}$ of the non-superconducting region
-- that is, the intermediate
layer and the half metallic region combined -- and the reflection 
matrix ${\cal R}_{\rm A}$ for Andreev
reflection off an ideal normal-metal--superconductor interface. Using
the same block structure as in Eq.\ (\ref{eq:Reeeh}), it
reads
\begin{equation}
{\cal R}_{\rm A}=\alpha(\varepsilon) 
\begin{pmatrix}
0 & i\sigma_2 e^{i \phi}\openone_{N_S}\\
-i\sigma_2 e^{-i \phi}\openone_{N_S} & 0
\end{pmatrix}
,\label{eq:RA}
\end{equation}
where $N_{\rm S}$ is the number of
propagating spin-degenerate orbital modes at the Fermi level at the 
superconductor interface and $\sigma_2$ is the Pauli matrix acting in
spin space, $\phi$ is the phase of the superconducting order
parameter, and
\begin{equation}
  \alpha(\varepsilon)=e^{-i \arccos (\varepsilon/\Delta)}.
  \label{eq:alpha}
\end{equation}
The scattering matrix ${\cal S}$ has the structure
\begin{equation}
{\cal S}=
\begin{pmatrix}
S(\varepsilon) & 0\\
0 & S(-\varepsilon)^*
\end{pmatrix},
\label{eq:Snorm}\end{equation}
where $S(\varepsilon)$ is the scattering matrix describing the
scattering of electron-like excitations off the non-superconducting region. 
The scattering matrix 
$S(\varepsilon)$ can be further divided into transmission and
reflection blocks, 
\begin{equation}
S=
\begin{pmatrix}
r & t'\\
t & r'
\end{pmatrix}
  \label{eq:Sblock}
,\end{equation}
where $r$ describes reflection for electrons coming from H, $r'$
describes reflection for electrons coming from the superconductor 
interface, and $t$ and $t'$ describe transmission from and to H. 
The matrices $r$ and $r'$ have dimension $N$ and 
$2 N_{\rm S}$, respectively. Solving for the total scattering matrix 
${\cal R}$ in terms of ${\cal R}_{\rm A}$ and ${\cal S}$, one then
finds
\begin{subequations}
  \label{eq:reereh}
\begin{align}
r_{\rm ee}&=r+\alpha^2 t' \sigma_2 r'^*\sigma_2(1-\alpha^2 r' \sigma_2 r'^* \sigma_2)^{-1}t\ ,\\
r_{\rm eh}&=ie^{i\phi}\alpha t'\sigma_2(1-\alpha^2 r'^*\sigma_2 r'\sigma_2)^{-1}t^*\ ,\\
r_{\rm he}&=-ie^{-i\phi}\alpha t'^*\sigma_2(1-\alpha^2 r'\sigma_2 r'^*\sigma_2)^{-1}t\ ,\label{eq:rhe}\\
r_{\rm hh}&=r^*+\alpha^2 t'^* \sigma_2 r'\sigma_2(1-\alpha^2 r'^* \sigma_2 r' \sigma_2)^{-1}t^*.
\end{align}
\end{subequations}
Here we suppressed the energy arguments; the complex conjugate
matrices in Eq.\ (\ref{eq:reereh})
should be taken at energy~$-\varepsilon$. 

In the scattering matrix approach, a necessary condition for the 
superconducting proximity effect is to have a nonvanishing $r_{\rm
  he}$. For an HS
junction, having a nonzero $r_{\rm he}$ is not automatic:
In the absence of spin-flip scattering in the intermediate 
layer, an electron coming from H is Andreev reflected as a spin-down 
hole. This cannot re-enter the half metallic contact; it is reflected
from the half metal instead, upon which it is Andreev reflected once more 
 to return as a spin-up electron.
Andreev reflection can occur only if the intermediate layer
is spin active, that is, its scattering matrix 
is {\em not}
diagonal in the spin up/down basis of the half-metallic contact. 
Such anomalous Andreev reflection, in which a spin-up electron coming 
from the half-metallic contact is reflected as a spin-up hole, 
is the key to the triplet
proximity effect.
Examples of spin active layers that make this possible are a 
ferromagnet with a magnetization direction not collinear with the 
polarization of the 
half metal, a normal metal with strong spin-orbit coupling, or a 
half-metallic spacer layer with a different polarization direction
and thin enough that there is nonzero transmission of minority 
electrons through evanescent modes.

In the next two sections we use the scattering theory to calculate the
conductance of an HS junction and the Josephson current in an SHS
junction.

\section{HS junctions}
\label{sec:3}

\subsection{General considerations}

\label{sec:Rsymm}

The scattering matrix ${\cal R}(\varepsilon)$ obeys particle-hole
symmetry,
\begin{equation}
  {\cal R}(\varepsilon)=\Sigma_{1}{\cal R}(-\varepsilon)^*\Sigma_{1},
\label{eq:ehsymmR}
\end{equation}
where $\Sigma_{1}$ is the first Pauli matrix acting in electron-hole 
space. For the special case $N=1$, this symmetry, in combination with
the condition that ${\cal R}(\varepsilon)$ is unitary, leads to the
condition that either $r_{\rm ee} = 0$ or $r_{\rm eh} = 0$ at the
Fermi level $\varepsilon=0$. As we show in the Appendix, generically 
one has $r_{\rm eh}(0)=0$, although the possibility $r_{\rm ee}(0)=0$
does occur for certain special choices of the spacer layer. The
case $N=1$ is relevant for the case that the contact to the half metal
has only one propagating mode at the Fermi level or, alternatively,
for the case that there is perfect translation symmetry in the
transverse direction so that different orbital modes do not mix.
To the best of our knowledge, the observation that Andreev
reflection at the Fermi level is absent for single-mode HS junctions has not been made
before. It presents a qualitative difference compared to FS junctions
in which both spin directions can propagate.

In the general theory of Sec.\ \ref{sec:2} the spin quantization axis
is taken to be the polarization direction of the half metal. Fixing
the spin polarization axis still allows for rotations around that
axis. For the scattering matrices appearing in the theory, such a
rotation is represented by the transformation
\begin{equation}
  S \to
  \begin{pmatrix}
  e^{i\psi/2} & 0 \\ 0 & e^{i \psi \sigma_3/2}
  \end{pmatrix}
  S
  \begin{pmatrix}
  e^{-i\psi/2} & 0 \\ 0 & e^{-i \psi \sigma_3/2}
  \end{pmatrix}, 
\end{equation}
where $S$ is the scattering matrix of the non-superconducting region, see
Eq.\ (\ref{eq:Snorm}), the block structure is that of Eq.\
(\ref{eq:Sblock}), and $\psi$ is the (azimuthal) angle of the 
rotation. Substituting this transformation into the expression
(\ref{eq:reereh}) for ${\cal R}$, one concludes that such a rotation
has the same effect on ${\cal R}$ as a change of the superconducting
order parameter $\phi$ as 
\begin{equation}
  \phi \to \phi + \psi.
  \label{eq:phitransform}
\end{equation}
A consequence of this observation is that, if the intermediate layer
is ferromagnetic or half metallic with a polarization along the unit vector
\begin{equation}
  \vm = (\sin \theta \cos \psi,\sin \theta \sin \psi,
  \cos \theta)^{\rm T},
\label{eq:magnvec}\end{equation}
which makes an angle $\theta$ with the polarization direction of the
half-metallic contact, ${\cal
  R}$ is a function of the difference $\phi-\psi$ only. 
(Here, and in what follows, the polarization of the half metal is taken 
to be along the $z$ axis.)
This
observation, which will be important in our discussion of the
Josephson effect in SHS junctions below,
was first made by Braude and Nazarov, using the
quasiclassical approach.\cite{Braude07} Here, it appears as a 
natural consequence of
the transformation rules of the scattering matrix under rotations.

\subsection{Ballistic HS junction with ferromagnetic spacer}
\label{sec:3b}

As a first and simplest application of the theory, we consider an HS
junction for which the intermediate layer is a ferromagnet. The
ferromagnet's magnetization points along the unit vector given 
in Eq.~\eqref{eq:magnvec}.
We take the interfaces on both sides of the ferromagnetic spacer 
layer F to be ideal and 
assume that the electron motion in F is ballistic. In 
that case, different orbital modes decouple, and one can use an 
effective single-mode description for each orbital mode $\mu$
separately. We also assume that the thickness of F is short in
comparison to the superconducting coherence length $\xi_S =
\hbar v_F/\Delta$ ($v_F$ is the Fermi velocity), so that the
energy-dependence of the scattering matrix $S$ can be neglected, and
we assume that the magnetic flux through F is small in comparison to
the flux quantum, so that the orbital motion is time-reversal
symmetric.

For this system, the calculation of $S$ requires the
composition of the $4 \times 4$ scattering matrix of the ballistic
ferromagnetic spacer layer,
\begin{equation}
  S_F = \begin{pmatrix}
  0& U\\
  U & 0
\end{pmatrix},\ \
  U=e^{i (\eta + \rho {\mathbf m}\cdot\bsigma)/2},
  \label{eq:SF}
\end{equation}
and the $3 \times 3$
scattering matrix $S_{H}$ of the ideal interface between the
half-metallic contact and the ferromagnetic spacer layer,
\begin{equation}
  S_H = \begin{pmatrix}
  0 & 1 & 0 \\
  1 & 0 & 0 \\
  0 & 0 & e^{i \beta}
  \end{pmatrix}.
  \label{eq:SH}
\end{equation}
In the above expressions, $\bsigma$ is the vector of Pauli matrices (acting in
spin space),  $\rho = \nu_{\uparrow} - \nu_{\downarrow}$ is the
difference of the 
phase shifts of majority and minority electrons in F upon
propagation through the spacer layer, and $\eta = \nu_{\uparrow} +
\nu_{\downarrow}$. In Eq.~\eqref{eq:SH}, $\beta$ is the phase shift
spin-down electrons experience upon reflection from the  half metallic
contact. The three phases $\rho$, $\eta$, and $\beta$ depend on the
orbital mode $\mu$. We have suppressed the mode dependence here, but
will restore it in the final expression, Eq.\ (\ref{eq:GEif}) below.
The block structure of $S_F$  is as in
Eq.~\eqref{eq:Sblock}. The same is true for $S_H$, where the lower 
right $2\times 2$ submatrix corresponds to the lower right block in
Eq.\ (\ref{eq:Sblock}).

Combining Eqs.\ (\ref{eq:SF}) and (\ref{eq:SH}) to calculate
$S$, and then using Eq.\ (\ref{eq:reereh}) to find ${\cal R}$, we obtain
\begin{widetext}
\begin{eqnarray}
  {\cal R}(\varepsilon) &=&
  \frac{\alpha^2}{1 + \alpha^2 \sin^2 \rho \sin^2 \theta}
%  \nonumber \\ && \mbox{} \times
  \begin{pmatrix}
    e^{-i \beta}(\cos \rho + i \sin \rho \cos \theta)^2 &
  -2 i (\varepsilon/\Delta) e^{i(\phi-\psi)} \sin \theta \sin \rho
  \\
  -2 i (\varepsilon/\Delta) e^{i(\psi-\phi)} \sin \theta \sin \rho
  &
  e^{i \beta}(\cos \rho - i \sin \rho \cos \theta)^2
  \end{pmatrix}.
  \label{eq:SFHif}
\end{eqnarray}
\end{widetext}
Substituting Eq.\ (\ref{eq:alpha}) for $\alpha$ and summing over all
orbital modes $\mu$, we conclude that
the differential conductance of a short ballistic HFS junction is
\begin{eqnarray}\label{eq:GEif}
  G(\varepsilon) &=& \sum_{\mu} 
  \frac{8 e^2}{h}  \\ && \nonumber \mbox{} \times
  \frac{\varepsilon^2 \sin^2 \theta \sin^2 \rho_{\mu}}
  {\Delta^2 (1 - \sin^2 \theta \sin^2 \rho_{\mu})^2 +
  4 \varepsilon^2 \sin^2 \theta \sin^2 \rho_{\mu}},
  \end{eqnarray}
where the summation is over the orbital modes in the HS junction.

This simple result illustrates the two main properties of the triplet
proximity effect in HS junctions: First, Andreev reflection is
possible as soon as there is a spacer layer that breaks spin-rotation
symmetry around the half-metal's polarization direction, 
provided the electron's spin precesses by an angle different from $0$
or $\pi$. [In Eq.\ \eqref{eq:GEif} this translates to the requirement
  that $\sin \theta \neq 0$ and $\sin \rho_{\mu} \neq 0$.] 
And, second, in the absence of orbital mode mixing, \mbox{$G = 0$} at 
the Fermi level, except for very
special choices of the thickness (proportional to $\rho_\mu$) and 
magnetization direction of the 
spacer layer. In the present case, these special choices are angles
$\theta$ and $\rho_{\mu}$ for which $\sin^2 \theta = \sin^2 \rho_{\mu} =
1$. In that case, one finds $G = (2 e^2/h) M$, where $M$ is the number 
of modes with $\sin^2 \rho_\mu=1$. 

\begin{figure}
\includegraphics[height=6cm]{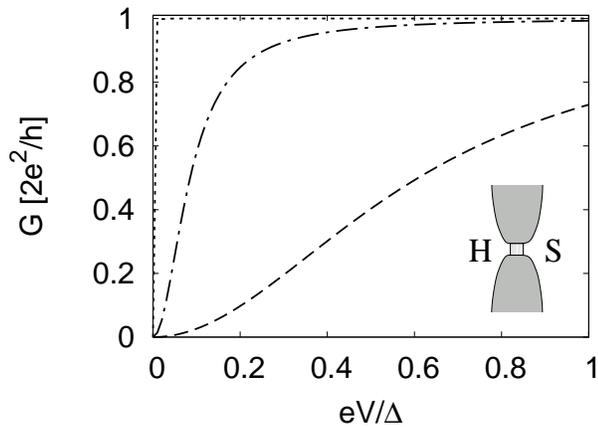}
\caption{The subgap differential conductance $G$
  versus the applied voltage $V$ for a ballistic
  single mode HS quantum point contact. 
  The small grey rectangle in the contact represents a region with
  a different magnetization than in the half-metallic part. Physically
  such a region can be present  due to a misaligned magnetization at
  the half-metal surface.\cite{eschrig2008tsc}
  In our calculations this  corresponds to the ferromagnetic spacer layer. 
  The   curves correspond to different values of the 
  phase angles in the ferromagnetic
  spacer, $\theta=0.8$ and $\rho=0.9$
  (dashed curve), $\theta=1.4$ and $\rho=1.2$, (dash-dotted curve), and $\theta=1.56$ 
  and $\rho=1.53$ (dotted curve).}
\label{fig:cond_qpc}
\end{figure}

Unlike the quasiclassical approach, the scattering approach can also
deal with systems in which the number of orbital modes is
small. The simplest way to illustrate this is to consider the contribution
of one orbital mode; in this case, the result in
Eq.~\eqref{eq:SFHif} and the corresponding term in Eq.~\eqref{eq:GEif}
describe a single mode ballistic ferromagnetic quantum point contact
between the half-metal and the superconductor. 
 In Fig.~\ref{fig:cond_qpc} we show the differential conductance  
of  such an HS quantum point contact for a few representative values of the
ferromagnet \mbox{parameters $\rho$ and $\theta$.} Both aforementioned
features are clearly seen: the conductance decreases as   $\sin^2
\theta$ and $\sin^2 \rho$ decreases, and it vanishes at the Fermi
energy.

\subsection{Ballistic HS junction with half-metallic spacer} 

If the spacer layer between the half-metallic reservoir and the
superconductor is not a ferromagnet, but a half metal, transmission
through the minority channel is via evanescent modes, not propagating
waves. The scattering matrix of the spacer layer, which was given by
Eq.\ (\ref{eq:SF}) for the case of a ferromagnetic spacer, now reads 
\begin{equation}
S_{H'} \, = \, e^{ - i \sigma_z \psi/2 } e^{ - i \sigma_y \theta/2}  S' 
 e^{ i \sigma_y \theta/2 } e^{ i \sigma_z \psi/2} ,
\end{equation}
where 
\begin{equation} 
  S'_{H'} \, = \, 
  \left( \begin{array}{cccc}
  0 & 0 & e^{ i \nu_{\uparrow} } &  0 \\
  0 &  - i e^{ i \nu_{\downarrow}} \sqrt{ 1 - \tau}  & 0 &  e^{ i \nu_{\downarrow}} \sqrt{\tau} \\
  e^{i \nu_{\uparrow} } &  0 & 0 & 0 \\
  0 &  e^{ i \nu_{\downarrow} } \sqrt{\tau} & 0 &  - i e^{ i \nu_{\downarrow}} \sqrt{ 1 - \tau}
         \end{array} \right). 
\end{equation}
The (mode-dependent) phase shift $\nu_{\downarrow}$ and transmission
coefficient $\tau$ for minority electrons are functions of the
wavefunction decay rate $q$ and effective mass $m_{\downarrow}$
of the evanescent minority electron
wavefunctions, the velocity $v$ of the majority electrons, and the
thickness $d$ of the half-metallic spacer layer. If $q d \gg 1$, the
minority electron phase shift $\nu_{\downarrow}$ becomes
independent of the layer thickness $d$,
\begin{equation}
  - i e^{i \nu_{\downarrow}} =
  e^{i \beta}
  = 
  \frac{v  - i \hbar q/m_{\downarrow} }{v +i \hbar q/m_{\downarrow} },
  \label{eq:nudown}
\end{equation}
whereas the transmission coefficient $\tau \propto e^{-2 q d}$
and $\nu_{\uparrow} = m_{\uparrow} v d/\hbar$, where $m_{\uparrow}$ is
the effective mass of majority electrons. [The phase shift $\beta$ is the
  reflection phase for minority electron reflection off a
  half-infinite half metal, see Eq.\ (\ref{eq:SH}) above.]

\begin{figure}
\epsfxsize=0.6\hsize
\epsffile{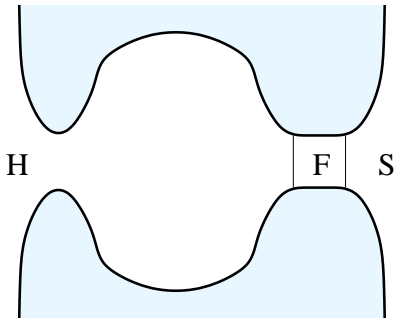}
\caption{\label{fig:3} (Color online) Chaotic HS junction, consisting of a
  half-metallic contact (left), a half-metallic quantum dot (center),
  a ferromagnetic spacer layer, and a superconducting contact
  (right).}
\end{figure}

With the definitions $\rho = \nu_{\uparrow} - \nu_{\downarrow}$ and
$\eta = \nu_{\uparrow} + \nu_{\downarrow}$, we then find that the
conductance of an HS junction with a half-metallic spacer is
\begin{equation} 
  \label{eq:GHH}
  G =  
  \frac{ 8 e^2}{ h} 
  \sum_{\mu}
  \frac{ \varepsilon^2 \Delta^2  \tau_{\mu} \left[ \sin  \rho_{\mu}  + (1 - \tau_{\mu})^{1/2}  \sin \eta_{\mu} \right]^2 \sin^2\theta }{ (B_0 \Delta^2 - B_1 \varepsilon^2 )^2 + 4 B_2^2 \varepsilon^2 ( \Delta^2  - \varepsilon^2 )},
\end{equation}
where we abbreviated 
\begin{widetext}
\begin{eqnarray*}
  B_0  & = & 
  [\sin \rho_{\mu} \cos \theta + (1-\tau_{\mu})^{1/2} \sin \eta_{\mu}]^2
  + [\cos \rho_{\mu} + (1-\tau_{\mu})^{1/2} \cos \eta_{\mu}]^2, \\
  B_1  & = & 
  2  + \left( 1 + \cos^2 \theta \right) ( 1 - \tau_{\mu})  
  + 2 (1 - \tau_{\mu})^{1/2} \cos( \eta_{\mu} -  \rho_{\mu}) (1 + \cos \theta), \\
  B_2  & = &
  1 + (1 - \tau_{\mu})^{1/2}  
  \cos( \eta_{\mu} -  \rho_{\mu}) (1+\cos \theta) 
  + ( 1 - \tau_{\mu}) \cos \theta,
\end{eqnarray*}
and restored the summation over the orbital modes $\mu$. For $\tau_{\mu}$
close to unity, this expression simplifies to the Andreev conductance
for an HS junction with a ferromagnetic spacer, Eq.\ (\ref{eq:GEif})
above. 
For small energies one may neglect the terms proportional to
$\varepsilon^2$ and $\varepsilon^4$ in the denominator, and we find
that $G \propto \varepsilon^2 \tau$. Since the transmission coefficients
$\tau_{\mu}$ are exponentially small if $q_{\mu} d \gg 1$, the 
conductance is dominated by the transverse mode $\mu$
with the lowest $q_{\mu}$. 

Similar to the case of ideal transmission, there is a special
set of parameters at which the conductance becomes large, independent 
of transmission. This occurs when the coefficient $B_0 = 0$
in Eq.\ (\ref{eq:GHH}), so that the denominator in that equation
vanishes at $\varepsilon=0$. The condition $B_0=0$ translates to
\begin{eqnarray}
  \cos \rho_{\mu} = -(1 - \tau_{\mu})^{1/2} \cos \eta_{\mu}, \nonumber
  \\
  \sin \rho_{\mu} \cos \theta = -(1 - \tau_{\mu})^{1/2} \sin \eta_{\mu}.
%\sin^2{\rho_{\mu}} \sin^2{\theta} & = & \tau_{\mu} 
%\\
%  (1 - \tau_{\mu})^{1/2} \cos{\eta_{\mu}}  +  \cos{\rho_{\mu}} & = & 0
%.
  \label{eq:conditions}
\end{eqnarray}
Solutions of Eq.\ (\ref{eq:conditions}) satisfy the relation $\sin^2
\rho_{\mu} \sin^2 \theta = \tau_{\mu}$, which generalizes the
condition for resonance
found for a ferromagnetic spacer layer (corresponding to
$\tau_{\mu}=1$). 
Since $\nu_{\downarrow\mu} =
  (\eta_{\mu} - \rho_{\mu})/2$
 is a material property if $q_{\mu} d \gg 1 $, see Eq.\ (\ref{eq:nudown})
 above, $\rho_{\mu} $
and $\eta_{\mu} $ are not independent in that limit. For a specific 
half metallic
material and in the limiting case $q_{\mu} d \gg 1$, the relevant
  solution of Eq.\ (\ref{eq:conditions}) then becomes
 $(\rho_{\mu} + \eta_{\mu})/2 = \nu_{\uparrow\mu} = \pi/2 \mod
  \pi$ and $\theta \to \pi$. Since $\nu_{\uparrow\mu}$ is a
  function of the thickness $d$ of the spacer layer, not a 
material property, this condition
  can always be satisfied for special values of $d$.
If a mode satisfies the conditions
(\ref{eq:conditions}), its contribution to the conductance is
\begin{equation}
  G_{{\rm res},\mu} 
  \, = \, 
  \frac{2 e^2}{ \hbar} 
  \frac{ 4 \Delta^2 \tau_{\mu}^2 }{4 \Delta^2 \tau_{\mu}^2 + 
  \varepsilon^2 \left\{ [ 1 - \cos{\theta} + \tau_{\mu} ( 1 +
  \cos{\theta})]^2 - 4 \tau_{\mu}^2 \right\}}.
\label{eq:gHHpres}
\end{equation}
At zero energy, one finds perfect Andreev reflection irrespective of
$\tau_{\mu}$. As before, the contribution of a single orbital mode in Eqs.~\eqref{eq:GHH}, \eqref{eq:gHHpres} describes 
the differential conductance of a single mode quantum point contact with a misaligned half metallic surface layer at the constriction, the analogue of the setup sketched in Fig.~\ref{fig:cond_qpc}.

\subsection{Chaotic HS junction}
\label{sec:qd}

As the next application of the scattering method, we consider 
a ``chaotic HS junction'', which consists of
a half-metallic contact, a chaotic quantum dot, a ferromagnetic
contact, and a superconductor, all connected in series, see
Fig.\ \ref{fig:3}. To illustrate the strengths of the scattering
matrix approach, we focus on a situation that is intractable with
quasiclassical methods: we   
restrict our discussion to the case that both contacts have 
one orbital mode only.

For definiteness, we take the quantum dot to be
half metallic, with the same polarization direction as the
half-metallic contact. Two alternative scenarios, a normal-metal
quantum dot and a ferromagnetic quantum dot with a magnetization
direction parallel to that of the half-metallic contact, will be
addressed at the end of this section. In all cases we assume that the typical
electron path length before exiting from the dot through one of the contacts is short
compared to the superconducting coherence length.

The calculation proceeds similar to that of the ballistic junction
shown above. For the chaotic HS junction with a half-metallic quantum
dot, we replace the scattering
matrix $S_H$ of Eq.\ (\ref{eq:SH}) by 
\begin{equation}
  S_{H} =
  \begin{pmatrix}
  - e^{i \chi} \sqrt{1 - \tau} &
  e^{i (\chi + \xi)/2} \sqrt{\tau} & 0 \\
  e^{i (\chi + \xi)/2} \sqrt{\tau} & 
  e^{i \xi} \sqrt{1 - \tau} & 0 \\
  0 & 0 & e^{i \beta}
  \end{pmatrix},
\end{equation}
where $0 \le \tau \le 1$ is the transmission coefficient of the 
quantum dot and $\xi$ and $\chi$ are scattering phases for 
reflection off the quantum dot. As before, $\beta$ is the phase shift
minority electrons acquire when they are reflected off the half metal.
The special case $\tau = 1$ simplifies to the ballistic HS point contact we
considered previously. For general $\tau$, one finds
\begin{equation}
G(\varepsilon)=\frac{2e^2}{h}
  \frac{16 \varepsilon^2 \Delta^2 
  \tau^2 \sin ^2 \rho \sin ^2\theta}{D(\varepsilon)}, 
  \label{eq:Ge}
\end{equation}
where
%\begin{widetext}
\begin{eqnarray}
  D&=&\left\{
  \Delta^2 (1-\tau)^{1/2}
  \left[
  \vphantom{M^M_M}
  4  \cos \theta \sin (\beta - \xi ) \sin (2 \rho)
%  \right. \right. \nonumber \\ && \left. \left. \mbox{} 
  +
  \cos (\beta - \xi) 
  \left(1-2 \cos (2 \theta)  \sin ^2 \rho +3 \cos (2 \rho) \right)
  \right]
  \nonumber \right. \\ && \left. \mbox{} + 2 (\tau
  -2) \left(2 \varepsilon^2+ \Delta^2 \sin ^2 \rho  \sin ^2\theta 
  - \Delta^2\right)
  \vphantom{()^{1/2}}\right\}^2
%  \nonumber \\ && \mbox{}
  +16 \varepsilon^2 \left(\Delta^2-\varepsilon^2\right) \tau^2.
\label{eq:Gexact}
\end{eqnarray}
\end{widetext}

For the special point $\sin^2 \theta = \sin^2 \rho = 1$ one has
$G(0) = 2 e^2/h$, independent of $\tau$ and the scattering phases
$\beta$, $\chi$, and $\xi$. The origin of this remarkable result is 
that, for $\sin^2 \theta = \sin^2 \rho = 1$, the
ferromagnet-superconductor interface not only provides 
perfect Andreev reflection between spin-up electrons and spin-up holes, 
but, moreover, after two subsequent Andreev
reflections the net phase shift is $-\alpha^2 = 1$ at the Fermi 
energy. Hence, combining the interface reflection matrix 
\eqref{eq:SFHif} with the scattering matrix of the quantum dot, the 
conductance at vanishing voltage is found to be
\begin{equation}
\frac{h}{2e^2}G(0)=\tau^2/[1-(1-\tau)]^2=1,
\end{equation}
 independent of the dot's transmission
coefficient $\tau$. This is to be contrasted to the corresponding formula
formula\cite{Bee92a}
\begin{equation}
\frac{h}{2e^2}G(0) = \tau^2/[1+(1-\tau)]^2
\end{equation}
 for the linear response conductance (per spin) of a superconductor in contact with a 
single mode quantum dot through a normal-metal contact (and without
magnetic a field). 
The difference arises from the fact that, in the 
latter case, the phase shift upon two Andreev reflections is 
$\alpha^2=-1$ at the Fermi energy.

\begin{figure}
\includegraphics[height=6cm]{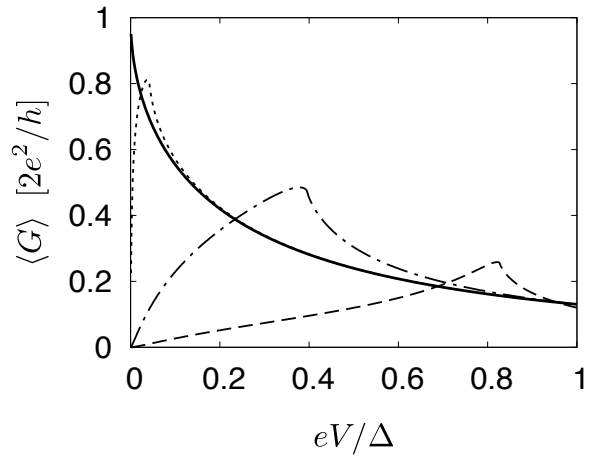}
\caption{The ensemble averaged subgap differential conductance $\langle G\rangle$
  versus the applied voltage $V$ for different
  phase angles $\theta$ and $\rho$ describing the ferromagnetic
  contact. The values of $\theta$ and $\rho$  are $0.8$ and $0.9$
  (dashed curve), $1.4$ and $1.2$, (dash-dotted curve), and $1.56$ 
  and $1.53$ (dotted curve), respectively. The solid curve shows
  the special case corresponding to $\theta=\rho=\pi/2$.}
\label{fig:cond_ch}
\end{figure}

For a chaotic quantum dot, the transmission coefficient $\tau$ and the
scattering phases $0 < \xi,\chi < 4 \pi$ are random quantities with the
statistical distribution\cite{RMTQTR}
\begin{equation}
  P(\tau,\xi,\chi) = \frac{1}{32 \pi^2} \tau^{-1/2}.
  \label{eq:Ptau}
\end{equation}
As is standard in the statistical approach to quantum transport,
the statistical ensemble is obtained by means of small variations of
the dot's shape or of the Fermi energy. In an experiment, both types 
of variations can be achieved by changing the voltage of nearby metal gates.
With the help of the distribution (\ref{eq:Ptau}) we can calculate the 
average Andreev conductance $\langle G \rangle$ for an 
ensemble of quantum dots. The phase shift $\rho$,  the angle
$\theta$ and the reflection phase $\beta$ are not averaged over, since they are properties of the
ferromagnetic contact and the half metal interface, not of the chaotic quantum dot.
The average can be performed analytically in the special case $\sin^2\rho = 
\sin^2 \theta = 1$, for which we find
\begin{equation}
\langle G\rangle =\frac{2e^2}{h}\left(
1-
  \frac{\Delta^2}{4 \varepsilon (\varepsilon^2-\Delta^2)^{1/2}}S\right)
  \label{eq:pipi}
\end{equation}
with
\begin{equation}
S=\sum_{\pm} 
 \pm x_{\pm}^3\\ \mbox{artanh} \frac{1}{x_{\pm}},
\end{equation}
where
\begin{equation}
 x_\pm^2 = \frac{2\varepsilon(\varepsilon \pm
  (\varepsilon^2-\Delta^2)^{1/2})}{\Delta^2}.
\end{equation}
 At the Fermi level $\langle G \rangle = 2e^2/h$,
in agreement with the discussion following Eq.\ (\ref{eq:Gexact}).
For general values of $\theta$ and $\rho$ no closed-form expression
for $\langle G \rangle$ could be obtained. The result of a numerical
evaluation of the ensemble average $\langle G(\varepsilon) \rangle$ 
is shown
in Fig.\ \ref{fig:cond_ch} for a few representative values of
$\theta$ and $\rho$. For generic $\theta$ and $\rho$, the
ensemble-averaged conductance vanishes at the Fermi level
$\varepsilon=0$. The quadratic dependence $G\sim \varepsilon^2$ for
$\varepsilon\rightarrow 0$ changes to a linear increase for relatively small
voltages. The conductance reaches a 
maximum at a voltage $e V$ below the superconducting gap $\Delta$ for
$\theta$,  $\rho$ sufficiently away from $\sin \theta =0$, $\sin \rho
=0$. With $\theta$,  $\rho$ approaching $\sin \theta =0$, $\sin \rho=0$, the 
position of the maximum moves towards the superconducting gap $\Delta$. For 
$\theta$,  $\rho$ close to $\sin \theta =0$, $\sin \rho=0$, the conductance is an
increasing function of the voltage in the full subgap regime.

For the case of a normal-metal quantum dot, the conductance
$G(\varepsilon)$ is given by Eq.\ (\ref{eq:Ge}), but with the
replacement $\beta - \xi \to \arctan[\tau \sin(\beta+\chi)/((2 - \tau)
  \cos(\beta+\chi) + 2 (1-\tau)^{1/2})]$. For the case of a
ferromagnetic quantum dot with a magnetization direction along that of
the polarization of the half-metallic contact, $G(\varepsilon)$ is
given by Eq.\ (\ref{eq:Ge}), but with $\beta-\xi$ replaced by a phase
shift $\beta'$ that is statistically independent of $\chi$, $\xi$, and
$\tau$. 
In both cases, the qualitative dependence of
  $\langle G \rangle$ on the parameters $\varepsilon$, $\theta$, and
  $\rho$ is the same as in the case of a half-metallic quantum dot
  discussed above.

\section{SHS junctions}
\label{sec:4}

We now contrast the transport current through an HS junction to the
supercurrent through an SHS junction. As in the previous section, we
consider the effect of a thin ferromagnetic layer between 
each superconductor and the adjacent half metal. (We do not consider
the case of a thin half-metallic spacer layer in this section.)
While, at zero temperature, the 
zero-bias conductance of a single single-channel HS
junction vanishes (except at special choices of the parameters), 
the zero temperature Josephson current $I$ is not zero.
The reason is that, in contrast to the linear response conductance
$G$, $I$ is not a Fermi level property. Instead, it is
determined by
the full excitation spectrum.

In order to apply the theory of
the previous sections, we consider the SHS junction as two opposing HS
junctions, see Fig.\ \ref{fig:2}. 
We refer to the opposing HS junction as S'H. 
Both junctions have intermediate ferromagnetic 
layers, which are denoted by F and F'. The two ferromagnets can
have different magnetizations, parameterized by polar angles
$\theta$, $\psi$ and $\theta'$, $\psi'$, respectively. The 
superconductors S and S' are assumed to have equal superconducting gaps
$\Delta$, but the phases $\phi$, $\phi'$ of the order parameters can 
differ.

Before turning to applications of our scattering theory, it is worthwhile to summarize some general considerations. 
Because of the transformation property (\ref{eq:phitransform}), the
Josephson current $I$ can depend on the superconducting phases $\phi$
and $\phi'$ and on the azimuthal angles $\psi$ and $\psi'$
through the single combination
\begin{equation}
  \tilde{\phi}=\phi-\phi'-(\psi-\psi')
\end{equation}
only. This observation was made previously in the context of the
 quasiclassical approach.\cite{Braude07,eschrig2007spc,eschrig2008tsc}

Under the operation of time reversal, the phases of the superconductors and the (position dependent) magnetization direction $\mathbf{m}$ 
transform as $\phi\rightarrow-\phi$, $\mathbf{m}\rightarrow -\mathbf{m}$. The supercurrent of the time reversed system is the opposite of the original, that is, 
\begin{equation}
I(\phi-\phi',\mathbf{m})=-I(\phi'-\phi,-\mathbf{m}).
\label{eq:ITR}\end{equation}
The supercurrent is invariant under a position independent rotation of the magnetization. This, together with Eq.~\eqref{eq:ITR} results in $I(\tilde \phi)=-I(-\tilde \phi)$.

For phase angles not close to the special
point $\sin^2 \theta =\sin^2 \theta' = \sin^2 \rho =\sin^2 \rho' =1$,
the Andreev reflection probability at the SH interfaces is significantly
smaller than unity [see Eq.\ (\ref{eq:SFHif}) above]. As a consequence, 
the $\tilde \phi$-dependence of the supercurrent is nearly sinusoidal in this
case. The detailed calculations of the next section show, however, 
that close to the special values of the phase angles  the  $\tilde \phi$-dependence becomes non-sinusoidal.

As an  illustration of our scattering theory, we now consider
 the ballistic and chaotic junctions addressed in the previous 
section. Our work on the Josephson effect in ballistic 
junctions complements that of
Galaktionov {\em et al.}, who used a Green function
approach.\cite{Galak08}

\subsection{Ballistic SHS junction}
\label{sec:BallisticSHS}

For the ballistic SHS junction different orbital modes are not mixed,
so that the scattering problem is effectively one-dimensional. As
before, we denote the difference of the (mode-dependent)
phase shifts of majority and
minority electrons transmitted through F by $\rho$, see Eq.\
(\ref{eq:SF}); The corresponding quantity for F' is denoted by
$\rho'$. We suppress the mode index $\mu$, except in the final
expressions.
For the calculation of the supercurrent, it is necessary 
that phase shifts accumulated inside the half metal are included into
the determinant in Eq.\ (\ref{eq:Ijosgen}). For an orbital mode $\mu$
these phase shifts depend on the length $L$ of the half-metallic 
segment and on the longitudinal component
$k_{\mu}(\varepsilon) = k_{\mu}(0) + \varepsilon/(\hbar v_{\mu})$ of the wave vector for that mode, where $v_\nu$ is the group velocity of the mode at $k_\mu(0)$.
In order to include
this  into Eq.\ (\ref{eq:Ijosgen}) 
we take the scattering matrix ${\cal R}'$ to include the scattering
phase shifts accumulated inside the half metal, 
\begin{eqnarray}
  {\cal R}' = \begin{pmatrix}
  e^{i k_\mu(\varepsilon) L} & 0 \\
  0 & e^{-i k_\mu(-\varepsilon) L} 
  \end{pmatrix}
  \tilde {\cal R}'
  \begin{pmatrix}
  e^{i k_\mu(\varepsilon) L} & 0 \\
  0 & e^{-i k_\mu(-\varepsilon) L} 
  \end{pmatrix},
  \nonumber \\
\end{eqnarray}
where $\tilde {\cal R}'$ is the reflection matrix for the S'H junction
without the scattering phases from the half metal. This matrix is
given in Eq.\ (\ref{eq:SFHif}) of the previous section, but with 
$\theta$, $\psi$, 
$\phi$, and $\rho$ replaced by $\theta'$, $\psi'$, $\phi'$, and
$\rho'$, respectively. 

Since there is a probability of normal reflection at each end of the
SHS junction, for a given orbital mode, the contribution to the supercurrent contains 
terms that oscillate with the length $L$ of the junction. 
For the total supercurrent, obtained by summing the contributions from
different orbital modes, however, this results only
in a small correction, provided that $k_{\mu}(0) L\gg 1$, since in this case, the  sum
of the oscillating contributions averages out.  
 Below, we calculate the non-oscillating
contribution to the Josephson current for a given orbital mode, and restrict our discussion
to the limiting cases of a ``short junction'' ($L \ll \xi_S$) and a ``long
junction'' ($L \gg \xi_S$). (In both cases, we assume that the
ferromagnetic spacer layers are thin in comparison to the
superconducting coherence length  $\xi_S$. The same assumption was
made in the previous section.)  

\begin{figure}
\epsfxsize=0.78\hsize
\epsffile{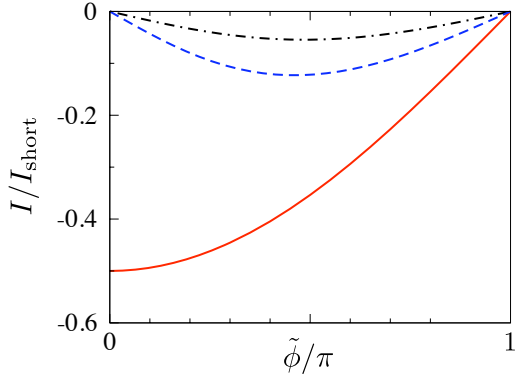}
\caption{\label{fig:scur} 
(Color online) The contribution of a single transverse
mode to the zero temperature supercurrent 
$I$ of a short SHS junction, as a function of 
$\tilde \phi$, for ferromagnetic phase angles  $\theta = \theta' = \rho = \rho' = \pi/2$ (solid), 
$\theta = \theta' = \rho = \rho' = \pi/4$ (dot-dash), and
$\theta = \theta' = \pi/2$, $\rho = \rho' = \pi/4$ (dashed). The
supercurrent is shown in units of $I_{\rm short} = e \Delta/\hbar$.}
\end{figure}

For a short junction, one may neglect the energy-dependence of the
wavenumber $k_\mu(\varepsilon)$ in the half metal.
A closed-form expression valid for arbitrary temperatures
could be obtained for the special case 
$\sin^2\theta = \sin^2\theta' = 1$ for a 
mode $\mu$ with $\sin^2 \rho_{\mu} = \sin^2 \rho'_{\mu} = 1$ only. 
The contribution $I_{\mu}$ to
the supercurrent of such a mode is
\begin{equation}
  I_{\mu} = 
%\frac{e}{\hbar}
%  \sum_{n} 
%  \frac{\Delta^2 k_B T \sin(\tilde \phi)}{\Delta^2(1 + \cos(\tilde
%  \phi) - \omega_n^2} \nonumber 
%  \nonumber \\ &=&
  -\frac{e \Delta}{2 \hbar} \cos \frac{\tilde \phi + s_{\mu} \pi}{2} 
  \tanh \left( \frac{\Delta}{2 k_B T}
  \sin \frac{\tilde \phi + s_{\mu} \pi}{2}\right),
  \label{eq:Ishortspecial}
\end{equation}
where $s_{\mu}$ is defined through the relation
\begin{equation}
  (-1)^{s_{\mu}} = \sin \rho_{\mu} \sin \rho'_{\mu}. 
  \label{eq:smu}
\end{equation}
The $\pi$ shift  in the current-phase relationship associated with 
$s_{\mu}$
originates in the properties of the interface reflection matrix
\eqref{eq:SFHif}: for this matrix,  the transformation $\rho\rightarrow
\rho+\pi$ is equivalent to $\phi\rightarrow \phi+\pi$. 

In the limit of high temperatures $k_BT\gg \Delta$, one can find a 
closed-form expression for arbitrary values of $\theta$, $\theta'$, $\rho$,
and $\rho'$. Upon summation over all orbital modes, one has
\begin{equation}
  I = - \sum_{\mu}
  \frac{e}{\hbar} \frac{\Delta^2}{8 k_B T}
  \sin \tilde \phi  \sin \rho_{\mu} \sin \rho'_{\mu} \sin \theta \sin \theta'.
\end{equation}
Note that although the angles $\rho_\mu$, $\rho'_\mu$ are mode dependent, for 
sufficiently thin spacer layers the mode dependence is weak enough that all
modes contribute to the total Josephson current with the same
sign. The supercurrent is reduced once the thickness of the spacer layers is large enough that $\rho_\mu$, $\rho'_\mu\gg 1$.

A numerical evaluation of the contributions to the zero-temperature supercurrent
is shown in Fig.\ \ref{fig:scur} for a few choices of the angles
$\theta$, $\theta'$, $\rho$, and $\rho'$. Although the discontinuity at
$\tilde \phi = s \pi$ is smeared for generic values of the phase
angles, the order of magnitude of the supercurrent is the same as at
the special point $\sin^2\theta = \sin^2\theta' = \sin^2 \rho = \sin^2
\rho' = 1$. This is in contrast to the Fermi-level conductance of an HS
junction, which was zero for generic phase angles and finite at the
special point. As discussed above,
the reason why the supercurrent has a different
behavior is that it is not a Fermi-level property but, instead,
depends on the entire excitation spectrum. For energies far away from
the Fermi level, the Andreev conductance is not qualitatively
different at the special point and elsewhere, see 
Fig.\ \ref{fig:cond_qpc}.

\begin{figure}
\epsfxsize=0.75\hsize
\epsffile{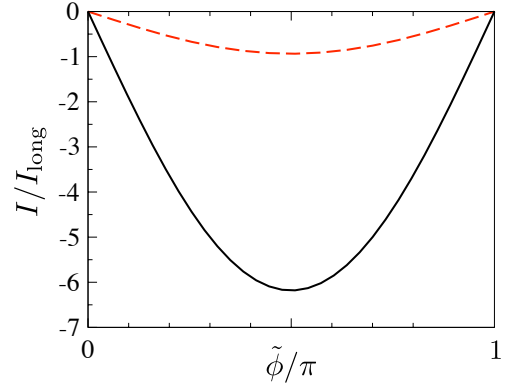}
\caption{\label{fig:scurlong} 
(Color online) The contribution of a single transverse
mode to the non-oscillating component of the zero temperature
supercurrent $I$ of a long SHS junction, as a function of 
$\tilde \phi$, for ferromagnetic phase angles  $\theta = \theta' = \rho=\rho' = \pi/3$ (solid) 
and $\theta = \theta' = \rho = \rho' = \pi/4$ (dashed). The current is
shown in units of $I_{\rm long} = e \hbar^2 v_{\mu}^3/\pi L^3
\Delta^2$, where $v_{\mu}$ is the mode-dependent longitudinal velocity.}
\end{figure}

For a long SHS junction (but still with ferromagnetic spacer layers
that are much thinner than $\xi_S$), again a
compact expression at arbitrary temperatures
could be obtained for the special case 
$\sin^2\theta = \sin^2\theta' =1$, for the contribution $I_{\mu}$ of a 
mode $\mu$ with $ \sin^2 \rho_{\mu} = \sin^2 \rho'_{\mu} = 1$ 
only. In this case one finds
\begin{eqnarray}
  I_{\mu} &=& - \frac{e}{\hbar} 2 k_B T \sum_{n}
  \frac{\sin (\tilde \phi + s_{\mu} \pi)}{\cosh(2 \omega_n L/\hbar v_{\mu}) - \cos
  (\tilde \phi + s_{\mu} \pi)},
  \label{eq:Ilongspecial}
\end{eqnarray}
where $s_{\mu}$ was defined in Eq.\ (\ref{eq:smu}).
At zero temperature the summation can be replaced by an integration
and one has
\begin{eqnarray}
  I_{\mu} &=& \frac{e v_{\mu} [\tilde \phi - ( 1 - s_{\mu}) \pi ]}{2 \pi L},\ \
  0 < \tilde \phi + s_{\mu} \pi < 2 \pi.
\label{eq:IT0spec}
\end{eqnarray}
In the limit of high temperatures, $T \gg \hbar
v_{\mu}/L$, only the term with $n=0$
contributes, so that
\begin{equation}
  I_{\mu} = - \frac{e}{\hbar} 4 k_B T e^{-2 \pi k_B T L/\hbar v_{\mu}}
  \sin (\tilde \phi + s_{\mu} \pi).
  \label{eq:IThighspecial}
\end{equation}
The special point $\sin^2\theta = \sin^2\theta' = \sin^2 \rho = \sin^2 \rho' =
1$ is singular, however, and the supercurrent contributions have a  qualitatively
different dependence on temperature for generic $\theta$, $\theta'$,
$\rho$, and $\rho'$. In the high-temperature regime $\hbar v_{\mu}/L \ll k_BT \ll \Delta$,  one finds
\begin{eqnarray}
  I &=& - \frac{e}{\hbar} \frac{16 \pi^2 k_B^3 T^3}{\Delta^2}
  \sum_{\mu}
  \sin \rho_{\mu} \sin \rho'_{\mu} \sin \theta \sin \theta'
  \sin \tilde \phi
  \nonumber \\ && \mbox{} \times
  \frac{e^{-2 \pi L k_B T/\hbar v_{\mu}}}{
  \left(1 - \sin^2 \rho_{\mu} \sin^2 \theta \right)
  \left(1 - \sin^2 \rho'_{\mu} \sin^2 \theta' \right)},
\end{eqnarray}
This result
is a factor $\sim(k_B T/\Delta)^2 \ll 1$ smaller (per orbital mode)
than the contribution for the
special choice of the angles $\theta$, $\theta'$, $\rho$, and $\rho'$
in Eq.\ (\ref{eq:IThighspecial}). Whereas the supercurrent of a short
Josephson junction depends on the full subgap excitation spectrum of the
junction,\cite{Bee92} the supercurrent in the long junction limit
is determined by the junction's excitation spectrum up to the Thouless
energy $\hbar v_F/L$ only.\cite{brouwer1997}
In this range of the spectrum, the absence
of Andreev reflection at the Fermi energy still strongly affects the
magnitude of the supercurrent. % if the junction is long. 
For temperatures below the Thouless energy $\hbar v_F/L$ the suppression
factor with which $I_\mu$ is reduced in comparison to the special case of 
Eq.\ (\ref{eq:Ilongspecial}) saturates around $(\hbar v_F/L \Delta)^2$.
No closed-form expressions for $I_\mu$ at arbitrary temperatures could be
obtained. Figure \ref{fig:scurlong}, shows $I_\mu$ versus $\tilde \phi$
at zero temperature, for two choices of the parameters $\theta$, $\theta'$, $\rho$, and
$\rho'$. 

\subsection{Chaotic SHS junction}

\begin{figure}
\epsfxsize=0.6\hsize
\epsffile{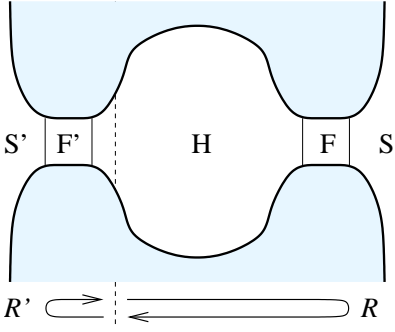}
\caption{\label{fig:5} 
  (Color online) Superconductor--half-metal-quantum-dot--superconductor junction. In
  the calculation, scattering from the half-metal quantum dot is
  included in the scattering matrix ${\cal R}$.}
\end{figure}

For a chaotic SHS junction, we include the quantum dot into ${\cal
  R}$, and take ${\cal R}' = \tilde {\cal R}'$ to be the scattering
  matrix of a junction without quantum dot, see Fig.\ \ref{fig:5}. 
For the chaotic SHS
  junction we only consider the limit that the superconducting coherence
  length is much longer that the typical electron path length in the dot before exiting
  through one of the contacts. 

\begin{figure}
\includegraphics[height=5.5cm]{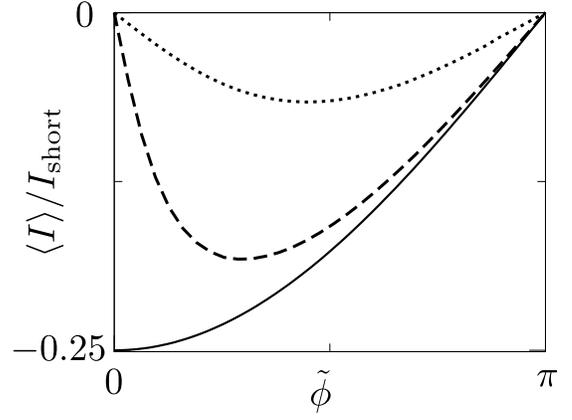}
\caption{The ensemble averaged Josephson current $\langle I\rangle$
      as the function of $\tilde{\phi}$ for different phase angles
  $\theta$, $\rho$, $\theta'$, $\rho'$ describing the ferromagnetic spacer layers. The values of
  $\theta$ and $\rho$ are $\theta=\rho=\pi/2$ (solid curve),  $0.9 \ \pi/2$ and
  $0.99\ \pi/2$ (dashed curve), and $0.5\ \pi/2$ and $0.6\ \pi/2$ (dotted curve),
  respectively. The values for the second contact are $\theta'=1.05\ \pi/2$
  and $\rho'=0.95\ \pi/2$. The
supercurrent is shown in units of $I_{\rm short} = e \Delta/\hbar$.
}
\label{fig:Jcurravg}
\end{figure}

In the special case $\sin^2\theta = \sin^2\theta' = \sin^2 \rho =
\sin^2 \rho' = 1$ the expression for the supercurrent is the same as
for a ballistic SHS junction, but with the replacement $\Delta \to
\Delta \tau^{1/2}$, where $\tau$ is the transmission coefficient of
the quantum dot. Since $\langle \tau^{1/2} \rangle = 1/2$, at 
zero temperature, the ensemble-averaged supercurrent is
\begin{equation}
  \langle I \rangle =
  -\frac{e \Delta}{4 \hbar}
  \cos \frac{\tilde \phi + s \pi}{2},
  \ \
  0 < \tilde \phi + s \pi < 2 \pi,
  \label{eq:Ichaoticshort}
\end{equation}
where $s$ was defined below Eq.\ (\ref{eq:Ishortspecial}).
No closed-form expressions could be obtained for generic values of
$\theta$, $\theta'$, $\rho$, and $\rho'$. A numerical evaluation of
the ensemble-averaged supercurrent is shown in Fig.\
\ref{fig:Jcurravg}. For phase angles close to the special case
discussed above, the supercurrent in a short
chaotic SHS junction follows Eq.\ (\ref{eq:Ichaoticshort}), except
near the discontinuity $\tilde \phi = s \pi$, which is smoothed out
away from the special point $\sin^2\theta = \sin^2\theta' = \sin^2 
\rho = \sin^2 \rho' = 1$. 
For phase angles not close to the special
point, the $\tilde \phi$-dependence of the ensemble-averaged
supercurrent is nearly sinusoidal, confirming the general observations made in
the beginning of this section.

\section{Conclusion}
\label{sec:5}
For the conventional proximity effect, the possibility of Andreev
reflection of electrons at the Fermi level gives a nonzero linear
conductance through a normal-metal--superconductor interface. In 
this article, we found that the situation is more delicate for the
triplet proximity effect in half-metal--superconductor (HS)
junctions. In the case that there is only one conducting channel at
the HS interface, or that different orbital channels at the HS
interface decouple, we found that Andreev reflection processes can 
be present only away from the Fermi level (except for special
choices of the interface parameters). 
While this result, which is independent of the nature of the spin 
active spacer layer in the HS junction, leads to a vanishing linear 
conductance, it allows for a nonzero Josephson current through 
an effectively single-channel SHS junction. We have illustrated this 
statement on systems both in the quasiclassical and in the
fully quantum mechanical regimes. In our calculations we have mainly 
concentrated on the case of ferromagnetic spin active intermediate layers.

First, we have calculated 
the zero temperature differential Andreev conductance at finite bias for short
HS junctions.
This is the observable in which the present Andreev reflection processes
manifest themselves in the most direct way.
Using the scattering matrix approach, we calculated the dependence of the
Andreev conductance on the phase angles of the spacer for all subgap voltages.
Our result in Eq.~\eqref{eq:Ge} can be used to describe the conductance of
a system with an arbitrary single channel structure in the half metal,
provided that its normal state scattering matrix is known. As an application,
we considered the case that the structure is a chaotic quantum dot and we
calculated  the ensemble averaged conductance from the known distribution of
the dot scattering matrices. In addition to the calculation of the differential conductance for systems with 
ferromagnetic spacer layer, we also studied ballistic systems where the spacer is a thin half metallic layer.

Second, we calculated the zero bias Josephson current through SHS junctions.
We have confirmed the observation, reported in earlier 
works,\cite{Braude07,eschrig2007spc,eschrig2008tsc} that the 
Josephson current depends on the
superconducting phase through the single variable
$\tilde{\phi}=\phi-\phi'-(\psi-\psi')$ only, which is
the difference of the superconductor phase difference of two
superconducting reservoirs and the azimuthal angle differences of the
magnetization direction of the two ferromagnetic spacer layers in the SHS 
junction. In the
framework of the scattering matrix approach, this observation follows
directly from the fact that the phase of the superconductor and the azimuthal
angle of the ferromagnetic spacer at an HS interface enter in
identical ways in the calculation of the Andreev reflection amplitude.
Further symmetry considerations showed that the supercurrent is an odd function of the variable $\tilde{\phi}$. 
Similarly to earlier works\cite{Braude07,Galak08,eschrig2007spc,eschrig2008tsc}, we also find that for
symmetric ferromagnetic spacers,  $\psi=\psi'$, $\rho=\rho'$,
$\theta=\theta'$, the sign of the current is the
opposite to the case of conventional SNS junctions [see Figs.~\ref{fig:scur},\ref{fig:scurlong},\ref{fig:Jcurravg}]. Consequently, the
equilibrium phase difference corresponds to $\phi-\phi'=\pi$,  i.e., a
$\pi$-junction behavior is realized. For independent configurations in
F and F', the
equilibrium phase difference varies continuously as the function of
interface parameters.

\begin{figure}
\includegraphics[height=5cm]{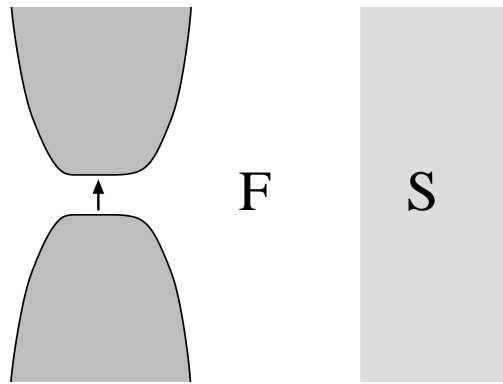}
\caption{Sketch of a possible experimental setup for testing the
  vanishing Andreev reflection at the Fermi level: a single channel
  quantum point contact to an FS junction. The arrow in the quantum point contact indicates that the point contact transmits only one
  spin direction.}
\label{fig:exp}
\end{figure}

It is worthwhile to compare our results for the Josephson current in single channel SHS systems
to the result for a single channel SNS systems.  In the latter case, at zero temperature and in  
the absence of magnetic field, for a perfectly transparent normal region, the (per spin) Josephson current is given by\cite{Bee92} \mbox{$I=(e\Delta/2\hbar) \sin(\phi/2)$}, for
short junctions and\cite{ishii1970jct} $I=ev_F\phi/2\pi L$ for long junctions, where $|\phi|<\pi$. 
We found  [see Eqs.\eqref{eq:IT0spec} and \eqref{eq:Ishortspecial}] that in the case of single channel SHS systems, in the special
point $\sin^2 \theta =\sin^2 \theta' = \sin^2 \rho =\sin^2 \rho' =1$, the
current-phase relation is identical,  apart from the phase shifts due to the azimuthal
angles and $s$. Away from the special point, the current-phase relation  becomes
sinusoidal, similar (apart from the phase shifts) to the case of a normal region with low transparency. 
By adjusting the interface parameters, 
the single mode triplet Josephson current interpolates between the result for
the conventional Josephson current through an ideal single mode channel
and through a tunnel barrier. The key property that distinguishes the
current phase relation in the triplet
Josephson effect through single mode structures from the conventional
Josephson effect is
the magnetization dependent phase shift. This is a feature that is common
between the fully quantum mechanical single channel limit and the multi mode case corresponding to the quasi-classics.

We end by relating our results about HS junctions to a possible
experiment. One experimental setup could be the HS quantum point
contact sketched in
Fig.~\ref{fig:cond_qpc}. Such a setup is somewhat subtle, as it relies on the presence of a surface
magnetization in the point contact. The generality of our
proof in the Appendix suggests,  however,  that the main features
of the single channel HS conductance, i.e., $G=0$ at Fermi level and $G\neq 0$ 
for $0<eV<\Delta$ could be tested in an experimentally more robust
arrangement. Such a setup could be a single channel point contact to an FS junction, as sketched
in Fig.~\ref{fig:exp}. It is not necessary to have the system in the
short junction limit, and  there can be arbitrary number of modes at the
ferromagnet-superconducting interface. The only important detail is that the junction ends in
a single mode point contact through which only one spin direction can
be transmitted. This can be achieved using a half metallic electrode or
with a spin filtering quantum point contact\cite{Potok02}.

\section*{ACKNOWLEDGMENTS}

We thank A. R. Akhmerov, H. Schomerus,  and I. Snyman for valuable discussions. 
This work was supported by the Dutch Science Foundation NWO/FOM, 
the Cornell Center for Materials research under NSF grant no.\ DMR 0520404,
the Packard Foundation, and by the NSF under grant no.\ DMR 0705476. 

\appendix
\section*{Appendix: Absence of Andreev reflection for single-mode HS junctions}
\label{app:proof}
In this appendix we prove that, generically, the Andreev reflection
amplitude $r_{\rm he}(0)=0$ for  a junction with $N=1$ orbital
modes in the half metallic side. The number of modes on the superconducting
side can be arbitrary. The starting point of the proof is the singular value
decomposition of the scattering matrix $S$ of the non-superconducting
region between the half-metallic and superconducting reservoirs,\cite{RMTQTR}
\begin{equation}
S=
\begin{pmatrix}
V & 0 \\
0 & W
\end{pmatrix}
\begin{pmatrix}
\hat{R} & \hat{T}^T \\
\hat{T} & -\hat{R'}
\end{pmatrix}
\begin{pmatrix}
V' & 0 \\
0 & W'
\end{pmatrix}.
\label{eq:Sdecomp}
\end{equation}
Here, $V$ and $V'$ are $N \times N$ unitary matrices, $W$ and $W'$
are unitary matrices of dimension $2N_S$, $N_{\rm S}$ being the number 
of orbital channels at the normal-metal--superconductor interface,
$\hat{T}$ is an $2 N_{\rm S} \times N$ matrix with
\begin{equation}
  \hat{T}_{kl}=\delta_{kl}\sqrt{\tau_l},\ \ k=1,\ldots,2N_S,\ \
  l=1,\ldots, N,
\end{equation}
with $\tau_l$ the $l$th transmission eigenvalue, $l=1,\ldots,N$, and
\begin{align}
\hat{R} =\sqrt{\openone_{N}-\hat{T}^T\hat{T}},\ \
\hat{R'}=\sqrt{\openone_{2N_S}-\hat{T}\hat{T}^T}.
\end{align}
Substituting the decomposition \eqref{eq:Sdecomp} in Eq.~\eqref{eq:rhe}, and
assuming $\det(\openone_{2N_S}+r'\sigma_2r'^*\sigma_2)\neq 0$, one finds 
\begin{equation}
r_{\rm he}(0)=-e^{-i\phi}V^*\hat{T}^T\left(Z^\dagger-\hat{R'}Z^*\hat{R'}\right)^{-1}\hat{T}V'
\label{eq:ehpolar}
\end{equation}
with $Z=W'^*\sigma_2 W$. If $N=1$, the
amplitude $r_{\rm he}(0)$ is proportional to the 11 element of the inverse in
\eqref{eq:ehpolar}. Using the general result $A^{-1}=(\det A)^{-1}{\rm
  adj}(A)$ for the matrix inverse, we find that this element is
proportional to the determinant of an antisymmetric matrix of dimension
$2N_S-1$, and is therefore zero. 
 The case $r_{\rm he}(0)\neq 0$ is possible if \mbox{$\det(\openone_{2N_S}+r'\sigma_2r'^*\sigma_2)= 0$}, 
that is, if the system has an Andreev bound state at $\varepsilon=0$ that is not coupled to the mode in
the half metal. For the ballistic HS system in Sec.~\ref{sec:3b}, 
\begin{equation}
  \det(\openone_{2N_S}+r'\sigma_2r'^*\sigma_2)=1-\sin^2 \rho \sin^2 \theta, 
\end{equation}
resulting in $\sin^2 \rho =\sin^2 \theta=1$ to be the only points where
$r_{\rm he}(0)$ can be nonzero.

\end{document}